\documentclass[aps,pra]{revtex4}

\usepackage{array}
\usepackage{amssymb,amsmath,amsbsy,amsgen,amsfonts}
\usepackage{latexsym}
\usepackage{graphicx}
\usepackage{amsfonts}
\usepackage{epsfig}
\usepackage{color}

\newcommand{\Tr}{\mathrm{Tr}}

\begin{document}

\title{Adaptive loop-based subtraction of a single photon from a traveling beam of light}
\author{
  Petr Marek,  Jan Provazn\'ik,   Radim Filip }
\affiliation{
   Department of Optics, Palack\'y University, 17. listopadu 1192/12, 77146 Olomouc, Czech Republic
}

\begin{abstract}
Manipulating light by adding and subtracting individual photons is a powerful approach with a principal drawback: the operations are fundamentally probabilistic and the probability is often small. This limits not only the fundamental scalability but also the number of operations that can be applied in realistic experimental settings. We propose and analyze an adaptive technique which can significantly increase the probability of success while preserving the quality of the photon subtraction. We show the improvement both in single mode preparation and manipulation of non-Gaussian states with negative Wigner functions  and in two-mode entanglement distillation protocol with Gaussian states of light.
\end{abstract}

\maketitle


\section{Introduction}
During the last decades, techniques manipulating individual quanta were used for testing many interesting and fundamental aspects of quantum physics \cite{Zavatta2004, Ourjoumtsev2006,Parigi2007}, in particular the applications in quantum communication where they allow entanglement distillation  \cite{Takahashi2010, entanglement_distillation2b} or noiseless amplification of traveling beams of light \cite{Usuga2010,Zavatta2010,Xiang2010,Donaldson2015,NPamplif}. Recently they have also expanded into the developing field of quantum thermodynamics \cite{thermo1,thermo2}. Overall, many quantum information protocols were and continuously are being tested in hybrid quantum optical experiments involving both discrete photons and continuous variable fields of light \cite{CVoptics4,CVoptics6,CVoptics7}.

The best approach towards manipulating the individual photons at optical frequencies has been, so far, achieved by employing the measurement-induced-operations paradigm. Projective measurements on states of atoms, which passed through cavity containing microwave field, were used to generate highly non-classical and non-Gaussian states of the field \cite{Delglise2008, Raimond2012}. Similar results were obtained in optical domain with detectors capable of distinguishing individual photons \cite{Lvovsky2002, Bimbard2010, Yukawa2013a,Cooper2013, Harder2016}. The detectors themselves can operate either on separate spatial modes, but they can also filter out particular frequency modes from the full spectrum of the light field \cite{Walschaers2017,Ra2017,Ansari2018}.

All of these implementations have one thing in common: the output states are always obtained by conditioning on one particular measurement result and, as a consequence, the operations are probabilistic in nature. This is a drawback both from the theoretical point of view, as it limits the protocols that can be meaningfully implemented, as well as from the experimental point of view. It is therefore quite important to work around this issue and to design operations which are deterministic or, if that cannot be done, which have as large probability of success as possible. It is worth mentioning here that the only operations that can be implemented deterministically are those emulating quantum dynamics described by positive trace preserving maps, such as \cite{Marek2011, Yukawa2013}. Some nonlinear operations, as is the case of the already mentioned photon subtraction and its compatriots, can never be performed deterministically. However, for the sake of efficient technologies and also the immediate experimental tests of noiseless amplification \cite{Usuga2010,Zavatta2010,Xiang2010,Donaldson2015,Park2016} or entanglement distillation \cite{Takahashi2010, entanglement_distillation2b}, it is important and worthwhile to improve the success rate of these operations beyond what is possible today.

In this paper we take a closer look at the photon subtraction operation and propose a way of increasing its probability of success while preserving its fundamental properties. The operation is probabilistic by definition but for sufficiently intense quantum state of light we can achieve success probabilities close to one, similarly to the methods employing atoms for detection \cite{Honer2011,Rosenblum2016}. We are, however, not limited to any class of states in particular. Our method employs standard avalanche photo diodes and is based on recycling the used quantum state after an unsuccessful photon subtraction and repeating the procedure until it either succeeds or until a pre-determined number of steps is achieved. This approach is adaptive; the results of the measurements shape the dynamics of the operation.
In \cite{Marshall2015} a similar method was briefly proposed but it was not analyzed in detail. In this paper we thoroughly discuss the cost for increasing the success rate, which is the decoherence of the state, and show how it depends on the parameters of the operation for both single mode and two-mode scenarios. We observe that the success probability can be significantly increased while the fundamental properties of the transformation are kept intact. The experimental realization of the proposed technique requires a quantum delay element for the traveling light in order to act on the light and to release it when needed. Fortunately, the recent experimental advances show a lot of promise in this area \cite{Yoshikawa2013, Takeda2017}.

\section{Single photon subtraction}
Subtraction of a single photon from a traveling mode of light is usually done by tapping off part of the light field on a strongly unbalanced beam splitter and subjecting this stray light to a detection by an avalanche photon detector (APD) that can only distinguish between photons and no photons. When the detector registers a photon, the operation is considered successful and the remaining state of light is kept, otherwise it is discarded. Formally, the detection event transforms an arbitrary quantum state $\rho_{0,in}$ into
\begin{equation}\label{}
    \hat{\rho}_{0, out} = \Tr_1 [ \hat{U}_{01} \hat{\rho}_{0,in}\otimes|0\rangle_1\langle 0| \hat{U}^{\dag}_{0,1} \hat{\Pi}_{1}^{\bullet}],
\end{equation}
where $\hat{U}_{01} = \exp[ \kappa (\hat{a}_0 \hat{a}_1^{\dag} - \hat{a}_0^{\dag} \hat{a}_1)]$ is the unitary operation of the beam splitter between the modes labeled by $0$ and $1$, with $\hat{a}_0$ and $\hat{a}_1$ being the corresponding annihilation operators for the two modes with the standard commutation relations $[\hat{a}_k,\hat{a}_j^{\dag}] = \delta_{kj}$. The two possible outcomes of the measurement by APD can then be represented by its Positive Operator Value Measure (POVM) elements, which are $\hat{\Pi}_{1}^{\circ} = |0\rangle_1\langle 0|$ and $\hat{\Pi}_{1}^{\bullet} = \hat{1} - |0\rangle_1\langle 0|$ for the ideal detector and $\hat{\Pi}_{1}^{\circ} = \sum_{n=0}^{\infty}(1-\eta)^n|n\rangle_1\langle n|$ and $\hat{\Pi}_{1}^{\bullet} = \hat{1} - \hat{\Pi}_{1}^{\circ}$ for a realistic detector with nonunit quantum efficiency~$\eta$. We neglect the dark counts as they are negligibly small for high quality superconducting detectors \cite{Jeannic2016}.
When the beam splitter is strongly unbalanced and the amplitude transmissivity $t = \cos \kappa$ is close to one, the unitary operator can be replaced by its Taylor expansion up to the first order and the output state can be approximated to
\begin{equation}\label{rhoout}
    \hat{\rho}_{0,out} \approx \kappa^2 \hat{a}_0 \hat{\rho}_{0,in} \hat{a}_0^{\dag}.
\end{equation}
We can see that in this limit, $\kappa \rightarrow 0$, the operation does indeed approach the perfect subtraction of a single photon, but the probability of success, given by the trace of (\ref{rhoout}), approaches zero as well.

\begin{figure}
  \begin{center}
  \includegraphics[width=0.6\linewidth]{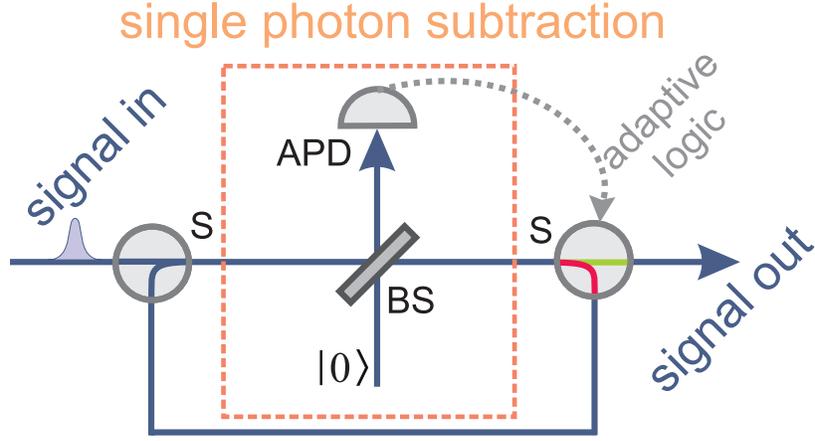}
  \end{center}
  \caption{
    (color online) Schematic representation of the adaptive photon subtraction. Part of the input signal is tapped off at the mostly transmissive beam splitter (BS) and directed towards avalanche photo-diode (APD). Positive detection event results in successfully transformed signal while negative one prompts the optical switches (S) to feed the signal back for the next attempt if a predetermined number of maximal steps was not reached yet.
  }
  \label{setup1}
\end{figure}

To increase this probability, we suggest modification of the scheme as per Fig.~\ref{setup1}. The main difference from the standard photon subtraction is that, when the detector measures no signal, we do not discard the quantum state, but instead loop it back to enter the beam splitter again. We repat this looping procedure until a photon is detected, after which we relase the field immediately. If a photon is not detected within a pre-determined number of loops $N$, we stop the procedure and discard the run. As a consequence the operation as a whole is still probabilistic. The transmission coefficient of the internal beam splitter is kept constant during the whole procedure. The architecture was inspired by the recent experimental proposals  \cite{Yoshikawa2013, Takeda2017}.
If the procedure succeeds in its $n$-th step we immediately stop the looping and the output state can be expressed as
\begin{equation}\label{rho_out_total}
    \hat{\rho}_{0,out}^{(n)} = \Tr_{1,\ldots,n}[ (\hat{U}_{(n)} \hat{\rho}_{0,in}\otimes \bigotimes_{k = 1}^{n}|0\rangle_k\langle 0|\hat{U}_{(n)})^{\dag} \hat{\Pi}_{(n)}],
\end{equation}
where
\begin{equation}\label{}
    \hat{U}_{(n)} = \bigotimes_{k=1}^{n} \hat{U}_{0,k}, \quad \hat{\Pi}_{(n)} = \bigotimes_{k=1}^{n-1}\hat{\Pi}_{k}^{\circ}\otimes\hat{\Pi}_{n}^{\bullet}
\end{equation}
are the collective operators of the sequences of the beam splitters and the projective measurements. The density matrix (\ref{rho_out_total}) is not normalized; its norm is equal to the probability of obtaining the positive detection event in that particular loop.  
If we do not care in which step the procedure succeeded, we need to average over the possibilities and the final output state can be expressed as
\begin{equation}\label{}
    \hat{\rho}_{0,out} = \frac{1}{P_S}\sum_{n=1}^{N} \hat{\rho}_{0,out}^{(n)},
\end{equation}
where $P_S = \Tr[\sum_{n=1}^{N} \hat{\rho}_{0,out}^{(n)}]$ is the overall probability of success. If we assume the individual beam splitter couplings to be weak we can approximate the output state as
\begin{equation}\label{rho_out}
    \hat{\rho}_{0,out} \approx \frac{1}{P_S}\hat{a}_0 \left[(1-t^2)\sum_{k = 0}^N t^{k \hat{n}_1} \hat{\rho}_{0,in} t^{k \hat{n}_1}  \right] \hat{a}_0^{\dag},
\end{equation}
with
\begin{equation}\label{success_probability}
    P_S \approx (1-t^2)\sum_{k = 0}^N \Tr[ \hat{\rho}_{0,in} t^{ 2k \hat{n}_0}\hat{n}_0],
\end{equation}
where $t = \cos(\kappa)$ is the transmission coefficient of the beam splitters. The probability of success depends on the transmission coefficient $t$ and on the maximal allowed number of steps $N$. For each input state $\rho_{0,in}$ and each value of $N$, it is possible to choose the coefficient $t$ in such way that any success rate $P_S$ from the interval $[0,P_{S,max}]$ can be obtained. Here $P_{S,max} = 1-\Tr[\rho_{in,0} \Pi_{1}^{\circ} ]$ is the probability of finding any photons in the state. In the upcoming analysis we will compare quality of the operation directly with the success rate, keeping the implicit dependence on $t$ hidden. However, the basic mechanism of improvement can be gleaned from (\ref{success_probability}), where we can see that if we fix the input state and the coefficient $t$ is sufficiently close to one, the probability of success increases with $N$ and for small values of $N$ the increase is almost linear. The cost of this improvement is the decoherence of the output state which is now a mixture a quantum states which underwent different noiseless attenuation processes \cite{conditeleport}. 
%
The output state is therefore a mixture, its decoherence being the cost for increasing the success rate.

\section{Analysis of the trade-offs}
We start with analyzing the adaptive loop-based approach for the archetypal example: subtraction of a single photon from a single mode pure squeezed state of light. In the ideal scenario this leads to creation of a squeezed single photon state. A distinctive property of this state is the negativity of its Wigner function, which also decisively marks the state as nonclassical and non-Gaussian \cite{negativity1, Mari2012}. One of the oftentimes seen consequences of decoherence is the reduction of this negativity. Its survival under decoherence was even proposed as one of the measures of nonclassicality \cite{nonclassicalitydepth}. So let us start by analyzing the behavior of the negativity at the center of the phase space, represented by the value $W(0,0)$ of the Wigner function, under the particular decoherence incurred by our operation.

\begin{figure}
  \begin{center}
    \includegraphics{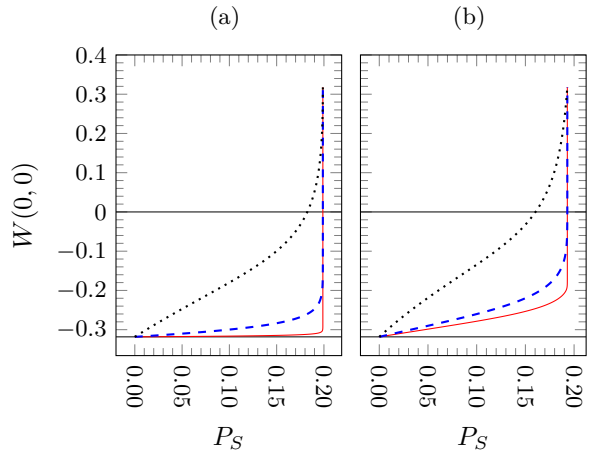}
  \end{center}
  \caption{
    Central negativity of non-Gaussian quantum state prepared from a pure state with $6$dB of squeeezing relative to the achievable probability of success for (a) ideal and (b) contemporary detectors with quantum efficiency $\eta = 0.8$. The performance of the original single step subtraction protocol (black dotted) is significantly surpassed by taking as little as 10 steps (dashed blue) and which can be further improved by performing 100 steps (solid red). While the improvement is significant for both the ideal and the realistic detection, the losses due to quantum inefficincy make it impossible to attain maximal negativity with maximal probability of success.
  }
  \label{fig.1}
\end{figure}

Let us first consider the ideal case of a pure squeezed state with $6$~dB of squeezing and $6$ dB of antisqueezing. In Fig.~\ref{fig.1} we can see the value of the Wigner function in relation to the maximal number of allowed steps $N$ and the probability of success $P_S$. For each particular state and a specific number of allowable steps $N$, the probability of success $P_S$ is in one-to-one correspondence with the transmissivity $t$ of the tapping mirror. For the case of the ideal detection we can see that for low numbers of allowed steps there is a distinctive trade-off between the achieved negativity and the probability of success. However, with increasing numbers of steps it becomes possible to approach maximal negativity at the maximal possible success rate and the improvement is significant already for ten steps. The maximal probability of success, $P_{S,max} = 1-\Tr[\rho_{in,0} \Pi_{1}^{\circ} ]$, depends solely on the input state and the efficiency of the detectors. For sufficiently intense signals the success rate approaches one.  However, unlike \cite{Honer2011,Rosenblum2016} who have shown this tendency for coherent states, our method is not limited for any particular class of states. In the case of currently available superconducting detectors with efficiency $\eta=0.8$ the maximal achievable negativity is lower; the reduced efficiency effectively acts as loss on the initial squeezed state and disrupts its parity properties. Consequently, the trade-off between the achievable negativity and the success probability never vanishes as in the ideal case. Nevertheless, for all values of success probability and negativity the improvement by considering multiple steps is clearly visible. The method can be therefore tested with already existing detectors.

\begin{figure}
  \begin{center}
    \includegraphics{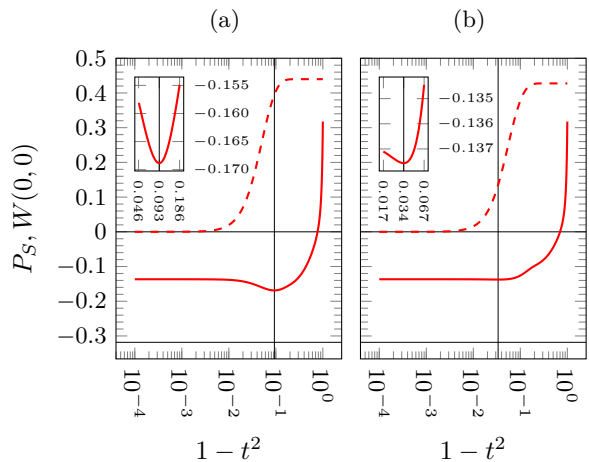}
  \end{center}
  \caption{
    The central negativity of the Wigner function (solid red line) and the achievable probability of success (dashed red line) in relation to the transmission coefficient of the tapping beam splitter for using the (a) ideal and (b) realistic superconducting detectors on an impure squeezed state with 8 dB of squeezing and 10 dB of anti-squeezing. The currently available superconducting detectors have quantum efficiency $\eta = 0.8$. The black vertical lines mark the points in which the Wigner function negativity reaches the local minima even though the success probability is not saturated. The insets show the detail of these points. The effect becomes observable for sufficiently squeezed states after a higher number of steps; the present figure is obtained for  N = 100 step procedure.
  }
  \label{fig.2}
\end{figure}

Let us comment on an interesting situation which can appear when the operation is applied to a squeezed vacuum state which is not pure. The presence of extra noise leads to larger values of $W(0,0)$. However, unlike the previous scenario, in which larger negativity was always associated with lower probability of success, in the case of impure squeezed states it is possible to simultaneously increase both the negativity and the probability of success. This behavior, which manifests for sufficiently efficient detectors and sufficiently high number of allowed steps, is illustrated in Fig.~\ref{fig.2} on the case of subtraction of photon from a squeezed state with  $8$ dB of squeezing and $10$ dB of anti-squeezing.

The negativity of Wigner function is an important quantum feature but it does not provide a complete picture of the decoherence. Specifically, when subtracting a photon from a squeezed state by the multi-step approach, the final quantum state is a mixture (\ref{rho_out}), but under the ideal conditions all the states in the mixture have Wigner function with maximal negativity at the origin. Let us therefore consider a different scenario. Instead of subtracting a photon from a squeezed state we will consider subtracting a photon from a superposition of two coherent states $|+\rangle \propto |\alpha\rangle + |-\alpha\rangle$. Such states can be used for quantum information processing protocols \cite{cats1,cats1b} and photon subtraction can be advantageously employed for probabilistic realization of quantum logic gates \cite{Marek2010, Tipsmark2011}. What limits the practical utilization of these states and, at the same time makes them an excellent `canary in a coal mine', is their fragility. When affected by loss or noise they tend to quickly lose coherence and deteriorate from a superposition of coherent states into their mixture, which has fidelity of only 0.5 with the initial superposition.
Although the fidelity is not a good quantifier of nonclassical interference aspects of the superposition, here it is a very sensitive measure of even small decoherence. Looking at the fidelity will therefore tell us more about the nature and the severity of the decoherence emerging during the adaptive operation.

\begin{figure}
  \begin{center}
    \includegraphics{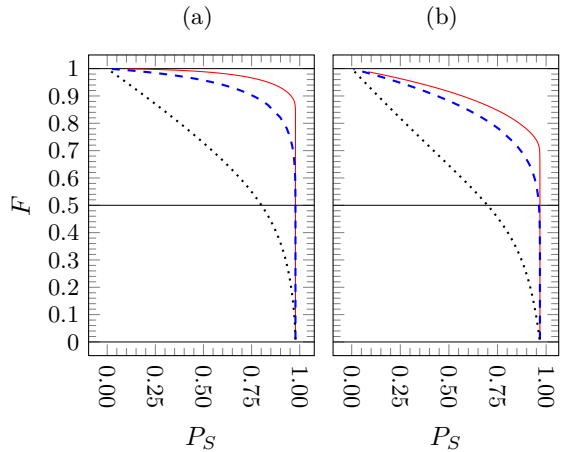}
  \end{center}
  \caption{
    Fidelity in respect to probability of success of the $\lvert + \rangle \mapsto \lvert - \rangle$ transition facilitated by the subtraction procedure employing (a) ideal and (b) realistic superconducting detectors with detection efficiency $\eta = 0.8$. The superpositions ${\lvert \pm \rangle \propto \lvert \alpha \rangle \pm \lvert - \alpha \rangle}$ are investigated for realistic $\alpha = 3/\sqrt{2}$ in both regimes of detection. Our procedure shows a significant improvement in success probability over the original subtraction protocol (dotted black) with as little as 10 steps (dashed blue). The maximal probability relative to desired fidelity of the transition can be further increased by taking, for example, 100 steps (solid red).
  }
  \label{fig.3}
\end{figure}

The ideal subtraction transforms the state $|+\rangle$ into a different superposition $|-\rangle \propto |\alpha\rangle - |-\alpha\rangle$, which is orthogonal to the original one.  As a figure of merit we can then consider the fidelity of the subtracted state $\hat{\rho}_{out}$ with the ideal state $|-\rangle$, which is $F = \langle - |\hat{\rho}_{out}|-\rangle$.
In Fig.~\ref{fig.3} we can see the dependence of the fidelity on the probability of success and the number of steps. When the employed detectors are ideal, even the fragile superposed coherent states can be transformed with both fidelity and the success probability achieving the maximal values, when the number of steps is large enough. At the same time, the improvement is significant already for ten steps. In the case of imperfect detectors the decoherence can not be avoided and the maximal fidelity is achievable only in the limit of $P_S \rightarrow 0$. However, already ten steps of the adaptive method can lead to obtaining the same fidelities with significantly higher probabilities of success.
It is therefore sufficient for immediate experimental test with high quality $|+\rangle$ superposition states
\cite{Huang2015,Jeannic2018}.

\section{Distillation of entanglement}
Photon subtraction is a crucial part of the entanglement distillation protocols \cite{entanglement_distillation1}. There it serves the dual role of increasing the entanglement and de-Gaussifying the quantum state \cite{entanglement_distillation2,entanglement_distillation2b} and even though it can be in principle substituted by other non-Gaussian operations such as photon addition \cite{Zavatta2004, Parigi2007}, photon subtraction is still the most feasible approach. For our purposes, entanglement distillation is also a suitable testing example, because the entanglement is vulnerable to decoherence differently than negativity of the Wigner functinon. 
We will show, however, that also in this scenario 	the improvements in success rate can overcome the costs.

\begin{figure}
  \begin{center}
    \includegraphics[width=0.6\linewidth]{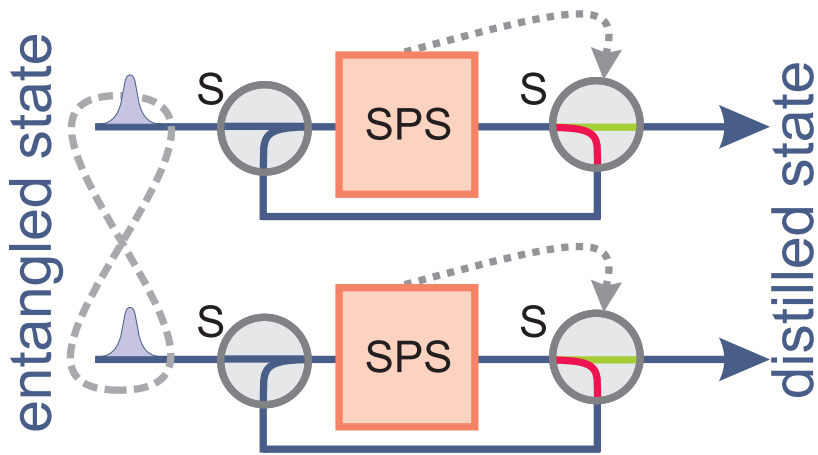}
  \end{center}
  \caption{
    Schematic representation of the distillation of entanglement with help of the adaptive photon subtraction. Single photon subtractions (SPS, depicted in detail in Fig.~\ref{setup1}) are attempted at both modes of the initial entangled state. If the subtraction was not successful and a predetermined number of steps was not reached yet, the  optical switches are used for feeding the signal back for the next attempt.
  }
  \label{setup2}
\end{figure}

In our test model we consider a pure two mode squeezed state
\begin{equation}
    \lvert \psi \rangle = \frac{1}{\cosh r} \sum_{f = 0}^{\infty}
        (\tanh r)^{f} \lvert f \rangle \lvert f \rangle
    .
\end{equation}
Such highly pure states are nearly available with almost $8$~dB of generalized squeezing. In order to increase the entanglement of the state we apply the adaptive photon subtraction in both of its arms, see Fig.~\ref{setup2}. We will consider only the first step of the distillation protocol, the degaussification. However, to reflect that we are ultimately interested in the properties of the Gaussian approximation of the final state we will quantify the entanglement of both the initial and the distilled state by Gaussian logarithmic negativity \cite{Adesso2007}, which can be calculated from the covariance matrix
\begin{equation}
  \sigma = \begin{pmatrix}
    \alpha & \gamma \\
    \gamma^{\intercal} & \beta
  \end{pmatrix}
\end{equation}
of the state using the PPT transformed symplectic invariant ${\Delta = \det \alpha + \det \beta - 2 \det \gamma}$ of the variance matrix $\sigma$ as $\mathcal{N} = \max \left\{ 0, - \log_{2} \nu \right\}$ with
\begin{equation}
  \nu = \sqrt{2 \left( \Delta - \sqrt{\Delta^2 - 4 \det \sigma} \right)}.
\end{equation}
\begin{figure}
  \begin{center}
    \includegraphics{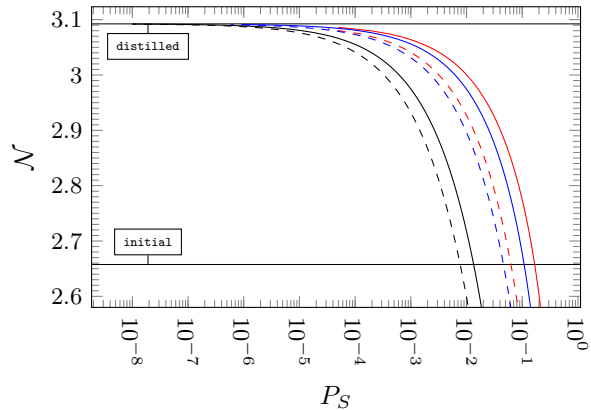}
  \end{center}
  \caption{
    Gaussaian logarithmic negativity in respect to success probability for distillation of entanglement from two mode uniformly squeezed vacuum state with $8$ dB of generalised squeezing. The behaviour is investigated for ideal (solid curves) and realistic superconducting (dashed curves) detectors with quantum efficiency $\eta = 0.8$. Our procedure shows a significant improvement in terms of probability for as little as $10$ allowed steps (blue). The attainable probability is roughly greater by an order of magnitude in comparison with the original procedure (black).
    Going to 75 steps (red) improves the probability even further, but the improvement rate steeply declines at that point.
  }
  \label{fig.4}
\end{figure}

In Fig.~\ref{fig.4} we show the dependence of the logarithmic negativity on the probability of success and the number of allowed steps. We have considered the symmetric scenario in which both modes of the entangled state were subjected to adaptive subtraction with the identical number of allowed steps. 
Overall we can say that the maximally achievable distilled Gaussian logarithmic negativity is preserved for the adaptive subtraction and that for all values of logarithmic negativity the adaptive procedure can reach higher success probability than the traditional single step technique. 
In all the cases, the maximal logarithmic negativity can be asymptotically reached in the limit  $P_{S} \rightarrow 0$, but the rate of convergence depends on the number of steps. For all distilled states with non-maximal logarithmic negativity the probability of obtaining them can be increased roughly by an order of magnitude when the adaptive subtraction with at most 10 steps is applied. Going to 75 steps increases the probability further, but our numerical calculations suggest the improvement begins to be saturated at that point. Interestingly enough, the qualitative improvement in success rate for all considered regimes is preserved even when the current superconducting detectors are considered.

\section{Conclusion and outlook}

We have proposed an adaptive protocol for improved photon subtraction which can achieve both higher probability of success and higher quality of the operation. The protocol is based on a simple premise: when a standard photon subtraction fails, that is when the triggering detector registers no event, the state itself is only minimally changed by noiseless attenuation. We can therefore recycle it and keep repeating the operation until it succeeds or until the state decoheres too much. We have analyzed the performance of the method for single mode photon subtraction from a squeezed and a superposed coherent states of light and for two-mode photon subtraction for use in distillation of entanglement. In all the cases, the procedure with ideal detectors allows obtaining the quality of simple single photon subtraction while significantly increasing the success probability, often up to its theoretical maximum. For the current superconducting detectors with nonunit quantum efficiency, the maximal probabilities can no longer be reached, but the improvement is still clearly visible. Remarkably, in both cases the improvements are significant already for the adaptive procedure with ten steps. Moreover, this regime might be naturally avoided by the upcoming highly efficient superconducting detectors \cite{SCdetectors, SCdetectors2}.

In this paper we have discussed only single photon subtraction, but the basic principle of the adaptive operation is not limited to it. In its place there could be any kind of probabilistic operation which is triggered by a certain measurement result. This could be as simple as multiple photon subtraction \cite{Usuga2010,Marek2010a}, but also operations such as photon addition \cite{Zavatta2004, Parigi2007} or also operations realized by coupling with discrete quantum systems \cite{Park2016}. Further extension lies in dynamical tuning of the interaction parameters. While we have considered identical couplings in all the loops of the operation, their dynamical adjustment would likely result in improving the trade-offs between the probability of success and the quality of the operation.

\section*{Funding}
P.M. and J. P. acknowledge grant GA18-21285S of the Czech Science Foundation. J. P. also acknowledges Palack\'y University project IGA PrF-2018-010.  R.F. acknowledges national funding from the MEYS and the funding from European Union's Horizon 2020 (2014-2020) research and innovation framework programme under grant agreement No 731473.

\bibliography{manuscript_submit}


%
%
%
%
%
%
%
%
%
%
%
%
%
%
%
%
%
%
%
%
%
%
%
%

\end{document}